\documentclass[conference]{IEEEtran}
\IEEEoverridecommandlockouts
\usepackage{cite}
\usepackage{amsmath,amssymb,amsfonts}
\usepackage{algorithmic}
\usepackage{graphicx}
\usepackage{textcomp}
\usepackage{xcolor}
\usepackage{url}
\usepackage[most]{tcolorbox}

\usepackage[caption=false,font=footnotesize]{subfig}

\def\BibTeX{{\rm B\kern-.05em{\sc i\kern-.025em b}\kern-.08em
    T\kern-.1667em\lower.7ex\hbox{E}\kern-.125emX}}
\begin{document}

\title{Real-Time Performance Benchmarking of TinyML Models in Embedded Systems (PICO: Performance of Inference, CPU, and Operations)\\
}

\author{
    Abhishek Dey,
    Saurabh Srivastava,
    Gaurav Singh,
    Robert G. Pettit\\
    adey6@gmu.edu,
    ssrivas6@gmu.edu,
    gsingh33@gmu.edu,
    rpettit@gmu.edu\\[1ex]
    Department of Computer Science, \\George Mason University, \\Fairfax, USA
}

\IEEEoverridecommandlockouts \IEEEpubid{\makebox[\columnwidth]{ 979-8-3315-9984-3/25/\$31.00~\copyright2025 IEEE\hfill} \hspace{\columnsep}\makebox[\columnwidth]{ }}

\maketitle



\begin{abstract}
This paper presents \textsc{PICO-tinyML-benchmark}, a modular and platform-agnostic framework for benchmarking the real-time performance of TinyML models on resource-constrained embedded systems. Evaluating key metrics such as inference latency, CPU utilization, memory efficiency, and prediction stability, the framework provides insights into computational trade-offs and platform-specific optimizations. We benchmark three representative TinyML models—Gesture Classification, Keyword Spotting, and MobileNet V2—on two widely adopted platforms, BeagleBone AI64 and Raspberry Pi 4, using real-world datasets. Results reveal critical trade-offs: the BeagleBone AI64 demonstrates consistent inference latency for AI-specific tasks, while the Raspberry Pi 4 excels in resource efficiency and cost-effectiveness. These findings offer actionable guidance for optimizing TinyML deployments, bridging the gap between theoretical advancements and practical applications in embedded systems.
\end{abstract}
\begin{IEEEkeywords}
TinyML, Embedded Systems, Real-Time Performance, Machine Learning, Benchmarking.
\end{IEEEkeywords}

\section{Introduction}
The rapid expansion of edge computing and the Internet of Things (IoT) \cite{madakam2015iotreview} has transformed data processing, enabling intelligent decision-making directly at the data source \cite{shi2016edgecomputing}. This paradigm shift has driven the adoption of TinyML \cite{warden2019tinyml}, a subset of machine learning focused on deploying lightweight models on resource-constrained devices like microcontrollers, wearables, and embedded systems. From gesture recognition in human-machine interfaces to real-time keyword spotting in voice-controlled devices, TinyML unlocks significant potential across diverse sectors, including healthcare, automotive, and industrial automation \cite{electronics13173562}.

However, deploying TinyML models on embedded platforms introduces unique challenges \cite{warden2019tinyml}. These devices are constrained by limited computational power, memory, and energy resources, making traditional machine learning models unsuitable for direct implementation. To address these limitations, techniques such as model quantization \cite{b1}, pruning \cite{han2024dtmmdeployingtinymlmodels}, and optimization \cite{le2023efficientneuralnetworkstiny} have been developed. While these techniques improve feasibility, they also introduce trade-offs, including reduced accuracy, increased inference latency, or higher resource consumption. These trade-offs emphasize the need for systematic benchmarking to assess model performance under real-world conditions.

Despite the growing interest in TinyML, existing tools and frameworks provide limited support for standardized, platform-agnostic benchmarking. Cloud-based solutions like Edge Impulse \cite{b2} focus on deployment but lack detailed insights into on-device inference metrics, particularly for resource-constrained environments. Furthermore, the heterogeneity of embedded platforms—ranging from general-purpose devices like Raspberry Pi \cite{raspberry_pi_4} to specialized hardware like BeagleBone \cite{beaglebone_ai64} AI64—adds complexity to evaluating model performance. Addressing this gap is critical for optimizing TinyML models for real-world applications.



This paper introduces \textsc{PICO-tinyML-benchmark}, a modular benchmarking framework designed to evaluate the real-time performance of TinyML models on embedded platforms. By systematically assessing key metrics such as \textbf{inference latency}, \textbf{CPU utilization}, \textbf{memory usage}, and \textbf{prediction confidence scores}, this framework provides actionable insights into the trade-offs between computational efficiency, resource constraints, and model accuracy. The benchmarking experiments focus on three representative TinyML models: gesture classification \cite{google_gesture_classification}, keyword detection \cite{mltk_kws_overview}, and MobileNet V2 \cite{pytorch_mobilenet_v2}—evaluated on two widely used platforms, BeagleBone AI64 and Raspberry Pi 4. The results offer a comprehensive perspective on platform-specific trade-offs, guiding developers and researchers in optimizing TinyML deployments across diverse application domains.

\textbf{1. Framework Design:} We developed \textit{\textsc{PICO-tinyML-benchmark}}, a versatile, platform-agnostic framework that benchmarks critical metrics, including inference latency, CPU utilization, memory usage, predicted labels, and confidence scores, providing actionable insights into TinyML model performance.

\textbf{2. Cross-Platform Benchmarking:} Using the framework, we systematically evaluated three models on two platforms:
\begin{itemize}
    \item \textbf{BeagleBone AI64:} Optimized for AI workloads, delivering consistent inference latency but with higher resource consumption.
    \item \textbf{Raspberry Pi 4:} A cost-effective platform demonstrating lower CPU and memory usage but with higher variability in latency.
\end{itemize}
\textbf{3. Performance Trade-offs:} Our experiments revealed key trade-offs:
\begin{itemize}

\item  \textbf{Latency vs. Resource Utilization:} Raspberry Pi 4 consistently outperformed BeagleBone AI64 in terms of inference latency, averaging \textit{1.76 ms} for gesture classification, compared to \textit{9.49 ms} on the BeagleBone AI64. However, the Raspberry Pi 4 consumed less CPU and memory on average.

\item  \textbf{Model Stability:} Both platforms exhibited high stability in prediction confidence scores, demonstrating their reliability for real-world applications.
    
\end{itemize}
\textbf{4. Practical Implications:} The insights gained from this study provide a roadmap for developers to optimize TinyML deployments based on application-specific requirements. For instance, latency-sensitive applications may prioritize platforms like BeagleBone AI64, while resource-constrained scenarios could favor Raspberry Pi 4.

\section{Related Works}
The growing field of \textit{TinyML} has spurred considerable interest in deploying machine learning models on resource-constrained devices. While advancements in model optimization techniques, such as quantization and pruning, have enabled the practical use of TinyML models, the evaluation of these models under real-world conditions remains a less explored area.

Several existing tools and frameworks focus on TinyML deployment. For instance, Edge Impulse \cite{b2} provides a cloud-based platform for building and deploying TinyML applications. However, its reliance on cloud resources and lack of fine-grained on-device benchmarking limits its applicability in evaluating real-time performance metrics such as inference latency, CPU usage, and memory utilization. Similarly, TensorFlow Lite Micro \cite{b3} supports deploying quantized models on embedded systems but does not include comprehensive benchmarking capabilities tailored for diverse hardware platforms.

Recent studies have explored the performance of TinyML models on specific hardware platforms. For example, Power-Performance Characterization\cite{tinyml_platform_comparison} compares the latency and power consumption of different TinyML models on microcontrollers. However, these studies are often limited to a single platform or a narrow set of metrics, making it challenging to generalize their findings to broader applications. Other works, such as MLPerf Tiny \cite{mlperf_tiny}, introduce benchmarking suites which focus on standardized benchmarks but primarily target inference accuracy rather than real-time resource utilization.

In contrast, this study introduces \textit{\textsc{PICO-tinyML-benchmark}}, a comprehensive and platform-agnostic framework designed to address the limitations of existing tools. Unlike cloud-centric platforms, our framework enables offline, on-device benchmarking, providing developers with actionable insights into real-time performance metrics. Furthermore, by evaluating three representative TinyML models across two widely used platforms—BeagleBone AI64 and Raspberry Pi 4—this study offers a broader perspective on the trade-offs between computational efficiency, resource constraints, and model stability.

To the best of our knowledge, this work is among the first to systematically benchmark TinyML models across multiple embedded platforms while integrating diverse metrics such as inference latency, CPU utilization, memory usage, and prediction confidence scores. This holistic approach bridges the gap between theoretical advancements in TinyML and their practical deployment, providing a foundation for optimizing TinyML applications in real-world environments.

\begin{figure}[tbp] 
    \centering
    \includegraphics[width=0.5\textwidth]{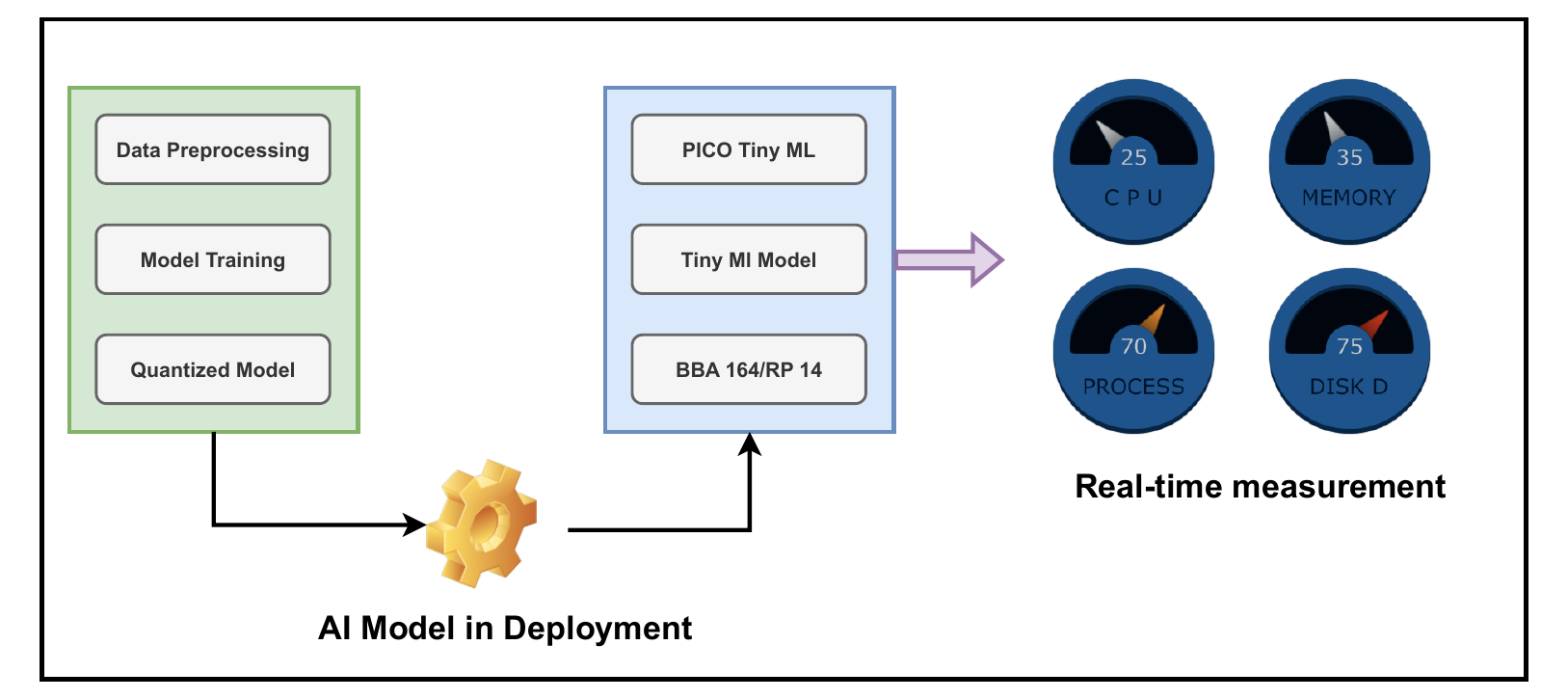} 
    \caption{Benchmarking process using PICO }
    \label{fig:example_label} 
\end{figure}
\section{Methodology}
This section presents the design and implementation of the \textsc{PICO-tinyML-benchmark} framework, a modular system for systematically evaluating the performance of TinyML models on resource-constrained embedded platforms. The benchmarking process is structured around four key metrics: inference latency, CPU utilization, memory usage, and prediction confidence scores. These metrics are collected through a detailed benchmarking procedure encompassing dataset preprocessing, inference execution, and metric logging across multiple iterations. Additionally, we detail the models, datasets, and hardware platforms utilized, followed by an overview of the tools and methods for visualization and statistical analysis. Finally, we discuss the real-world implications of the insights gained from these evaluations.

\subsection{Evaluation Metrics}



We evaluate model performance using four key metrics: \textbf{inference latency}, \textbf{CPU utilization}, \textbf{memory usage}, and \textbf{prediction confidence scores}. These metrics reflect real-time responsiveness, computational load, memory efficiency, and prediction reliability—essential indicators for deploying TinyML models in embedded systems. Detailed metric definitions and logging procedures are provided in Appendix~\ref{appendix:metrics}.

\subsection{Overview of Benchmarking Framework}

The \textit{\textsc{PICO-tinyML-benchmark}} framework is designed to enable systematic evaluation of TinyML models on embedded systems. It benchmarks the key metrics described above using tools such as TensorFlow Lite Runtime \cite{tensorflow_lite} for inference execution and \texttt{psutil} for real-time system monitoring. The framework follows a modular design, enabling seamless integration of models, datasets, and hardware platforms. The benchmarking process consists of four stages:
\begin{enumerate}
    \item \textbf{Dataset Preprocessing:} Preparing the input data to align with the model’s requirements, including resizing, normalization, and spectrogram generation.
    \item \textbf{Inference Execution:} Running inference on 100 iterations for each model on each hardware platform.
    \item \textbf{Metric Collection:} Monitoring and logging latency, CPU utilization, memory usage, and model outputs during each iteration.
    \item \textbf{Visualization and Analysis:} Generating visualizations to uncover performance trends and performing statistical analysis to evaluate trade-offs.
\end{enumerate}

This structured approach ensures consistency and reproducibility, enabling meaningful comparisons across platforms and models.

\subsection{Models and Datasets}

To comprehensively evaluate the benchmarking framework, we selected three representative models spanning diverse TinyML use cases: gesture recognition, audio keyword spotting, and image classification. Each model was paired with a curated dataset, ensuring compatibility with real-world applications.

The first model, \textbf{Gesture Classification}, is a quantized TensorFlow Lite model trained to recognize hand gestures from images. The dataset comprises 100 RGB images sampled from a public hand gesture dataset \cite{hand_gesture_dataset}, which were resized and normalized to match the model’s input dimensions while retaining gesture-specific features.

The second model, \textbf{MLTK Keyword Spotting}, is a pretrained and quantized TensorFlow Lite model for detecting spoken keywords from audio samples \cite{mltk_dataset}. The dataset includes 100 audio clips from the MLTK framework’s test set, preprocessed into spectrogram representations to align with the model’s input requirements.

The third model, \textbf{MobileNet V2}, is a quantized version of the MobileNet architecture tailored for image classification in TinyML environments \cite{mobilenet_v2}. A dataset of 100 web-sourced images was curated, resized, and normalized to fit the model’s input specifications. The compact architecture and efficiency of MobileNet V2 make it suitable for edge tasks like object detection and scene understanding.

By selecting models from distinct application domains and preparing datasets to reflect real-world scenarios, this study ensures the relevance and applicability of its benchmarking results.




\subsection{Hardware Platforms}

We evaluated the performance of the models on two widely adopted embedded platforms: the \textbf{BeagleBone AI64} and the \textbf{Raspberry Pi 4}. These platforms were chosen to represent diverse trade-offs in computational capability, cost, and optimization for machine learning workloads.

The \textbf{BeagleBone AI64} includes a Cortex-A72 CPU at 2.0 GHz, 8 GB of LPDDR4 RAM, and AI-specific hardware such as a PowerVR GPU and co-processors. It runs Debian-based Linux distributions and is commonly used in performance-focused edge applications.

The \textbf{Raspberry Pi 4} features a 1.5 GHz quad-core Cortex-A72 CPU and up to 4 GB of LPDDR4 RAM. Its accessibility and ecosystem support make it popular in general-purpose and energy-efficient TinyML deployments.

Detailed hardware specifications can be found at the official BeagleBone AI64 and Raspberry Pi 4 sites(\cite{beaglebone_ai64}~and~\cite{raspberry_pi_4}).

\subsection{Execution Configuration}

All models were executed using identical TensorFlow Lite configurations and were quantized for efficient inference on embedded platforms. To ensure a fair comparison, hardware accelerators on the BeagleBone AI64 were \textbf{disabled}, and all experiments were run using CPU-only execution on both platforms. Evaluating accelerator-assisted execution is left as future work to explore performance gains from AI-specific hardware such as GPUs or NPUs.

\subsection{Visualization and Statistical Analysis}

To uncover performance trends and trade-offs, the framework generates visualizations for each metric across models and platforms:
\begin{itemize}
    \item Distribution of inference latency over iterations.
    \item CPU and memory utilization trends.
    \item Prediction distributions and confidence score trends.
\end{itemize}

Statistical analysis complements the visualizations by summarizing key trends. Comparative studies between platforms highlight differences in performance, enabling insights into platform-specific strengths and weaknesses.

\begin{figure*}[h]
\centering
\subfloat[Gesture Classification (BBAI64)]{\includegraphics[width=0.32\textwidth]{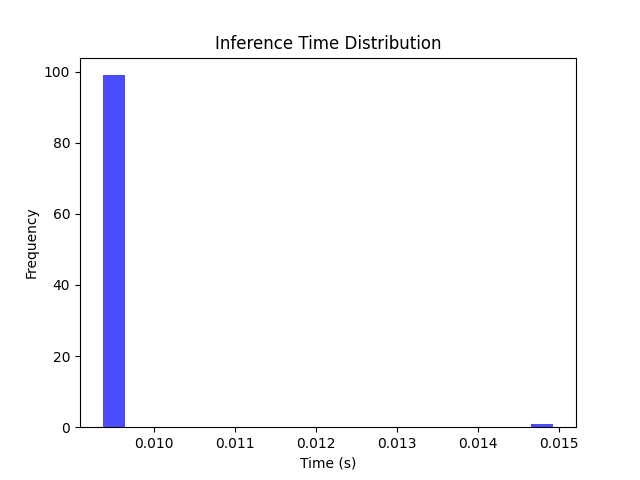}\label{fig:gesture_latency_bbai64}}
\hspace{-0.1cm}
\subfloat[Gesture Classification (RPI4)]{\includegraphics[width=0.32\textwidth]{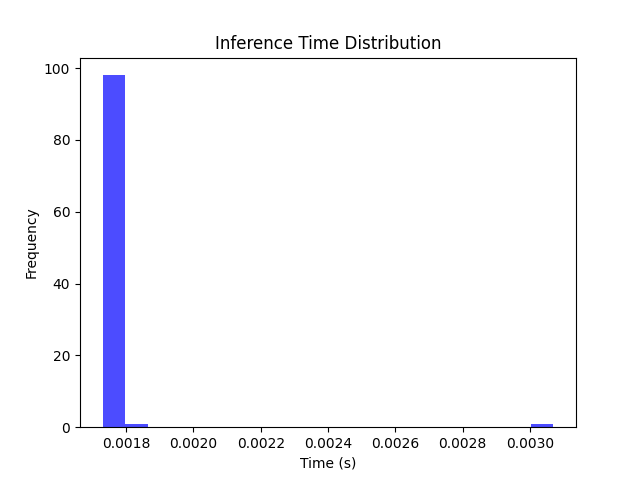}\label{fig:gesture_latency_rpi4}}
\hspace{-0.1cm}
\subfloat[Keyword Spotting (BBAI64)]{\includegraphics[width=0.32\textwidth]{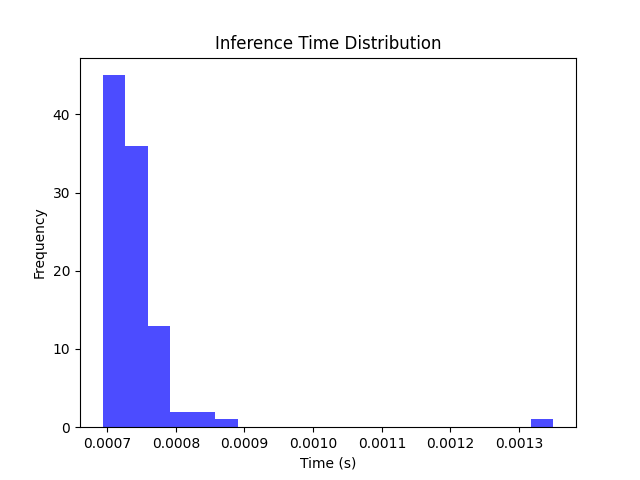}\label{fig:keyword_latency_bbai64}} \\
\vspace{-0.2cm}
\subfloat[Keyword Spotting (RPI4)]{\includegraphics[width=0.32\textwidth]{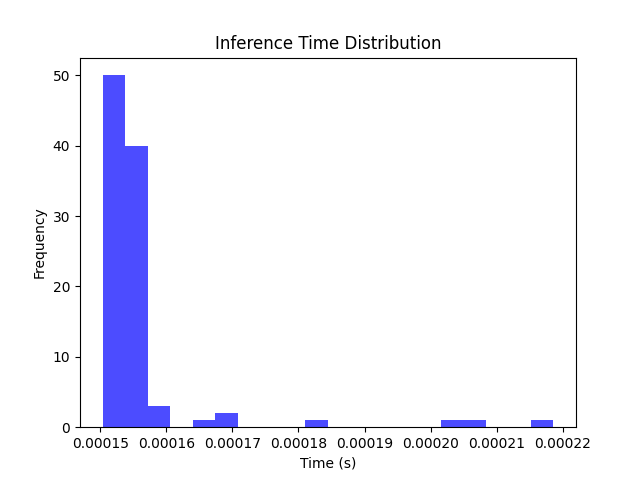}\label{fig:keyword_latency_rpi4}}
\hspace{-0.1cm}
\subfloat[MobileNet V2 (BBAI64)]{\includegraphics[width=0.32\textwidth]{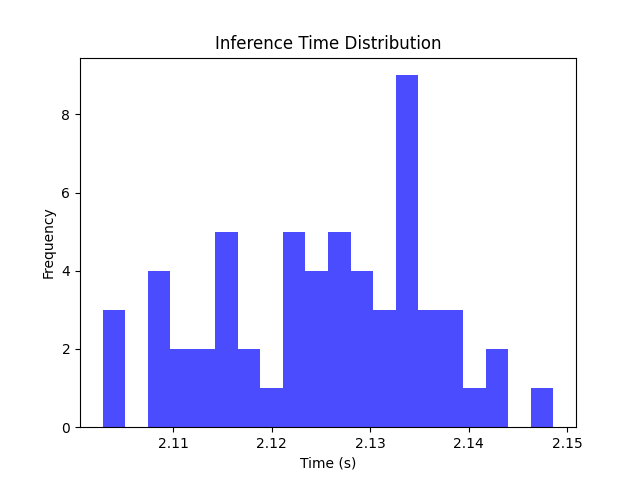}\label{fig:mn_latency_bbai64}}
\hspace{-0.1cm}
\subfloat[MobileNet V2 (RPI4)]{\includegraphics[width=0.32\textwidth]{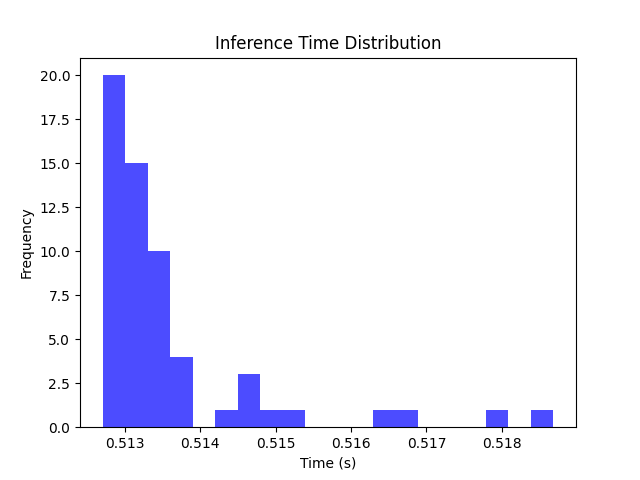}\label{fig:mn_latency_rpi4}}
\vspace{-0.1cm}
\caption{Latency Distribution for All Models Across BeagleBone AI64 and Raspberry Pi 4.}
\label{fig:latency_comparison}
\end{figure*}

\subsection{Real-World Applicability}

The benchmarking results have practical implications for deploying TinyML models in real-world environments:
\begin{itemize}
    \item \textbf{Latency Thresholds:} Results guide model selection for latency-sensitive applications, such as gesture recognition.
    \item \textbf{Resource Trends:} CPU and memory usage trends inform platform suitability for specific use cases and constraints.
    \item \textbf{Stability Analysis:} Confidence score trends identify potential inconsistencies in model performance, critical for reliable deployments.
\end{itemize}

By bridging the gap between theoretical model optimization and practical deployment, the proposed framework empowers developers to make informed decisions for resource-constrained TinyML applications.

\section{Results and Analysis}
This section evaluates the performance of Gesture Classification, Keyword Spotting, and MobileNet V2 models on BeagleBone AI64 and Raspberry Pi 4. The results address three key research questions (RQs), focusing on inference latency, resource utilization, and prediction stability.

\begin{table}[h]
\centering
\caption{Average Inference Latency (ms) Across Platforms}
\begin{tabular}{|l|c|c|}
\hline
\textbf{Model}         & \textbf{BeagleBone AI64} & \textbf{Raspberry Pi 4} \\ \hline
Gesture Classification & 9.49                    & \textit{\textbf{1.76} }                 \\ \hline
Keyword Spotting       & 0.74                  & \textit{\textbf{0.16}  }                \\ \hline
MobileNet V2           & 2125.04                  & \textit{\textbf{513.60 } }                \\ \hline
\end{tabular}
\label{tab:latency_comparison}
\end{table}

\textbf{RQ1: How do the platforms compare in terms of inference latency?}
Latency is a critical factor in real-time TinyML deployments. 
Table~\ref{tab:latency_comparison} presents the average inference latency for all models, while Figure~\ref{fig:latency_comparison} illustrates the distribution across platforms. \textit{Across all models, the \textit{Raspberry Pi 4} consistently outperformed the \textit{BeagleBone AI64} in latency.}

\begin{table}[h]
\centering
\caption{Resource Utilization Across Platforms}
\begin{tabular}{|l|c|c|c|}
\hline
\textbf{Model}         & \textbf{Platform}     & \textbf{CPU (\%)} & \textbf{Memory (\%)} \\ \hline
Gesture Classification & BeagleBone AI64       & 40.38              & 15.10                \\ 
                       & Raspberry Pi 4        & \textbf{\textit{8.88}}             & \textbf{\textit{11.00 }}               \\ \hline
Keyword Spotting       & BeagleBone AI64       & 18.50             & 18.96                \\ 
                       & Raspberry Pi 4        & \textbf{\textit{5.00 }}            & \textbf{\textit{11.00}}                \\ \hline
MobileNet V2           & BeagleBone AI64       & 51.80             & 19.75                \\ 
                       & Raspberry Pi 4        & \textbf{\textit{28.59}}             & \textbf{\textit{11.81}}                \\ \hline
\end{tabular}
\label{tab:resource_utilization}
\end{table}

\begin{figure*}[h]
\centering
\subfloat[CPU Usage for Gesture Classification (BBAI64)]{\includegraphics[width=0.3\textwidth]{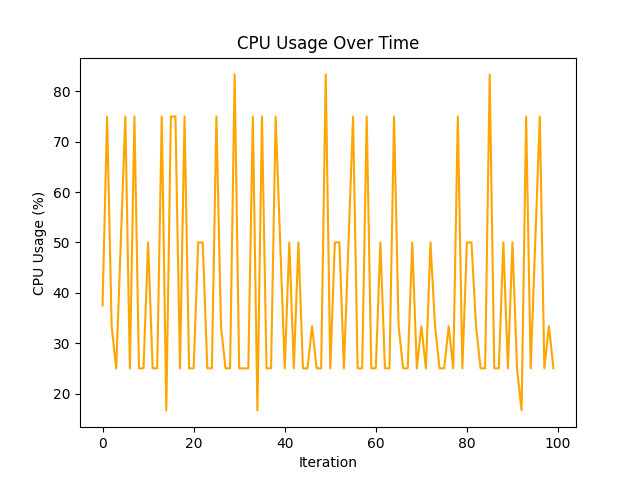}\label{fig:gesture_cpu_bbai64}}
\hspace{-0.1cm}
\subfloat[CPU Usage for Gesture Classification (RPI4)]{\includegraphics[width=0.3\textwidth]{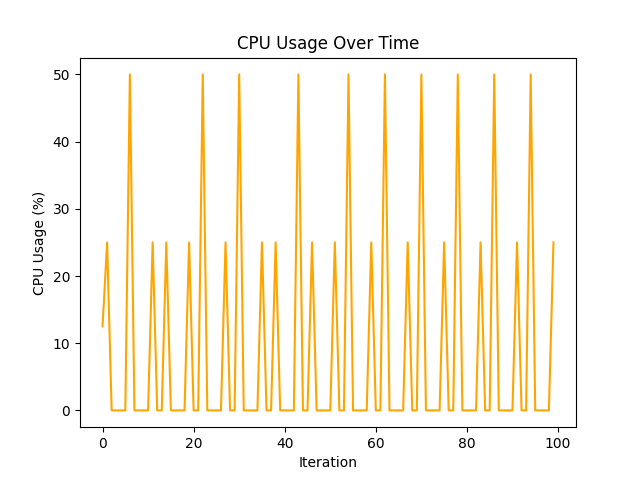}\label{fig:gesture_cpu_rpi4}}
\hspace{-0.1cm}
\subfloat[Memory Usage for Gesture Classification (BBAI64)]{\includegraphics[width=0.3\textwidth]{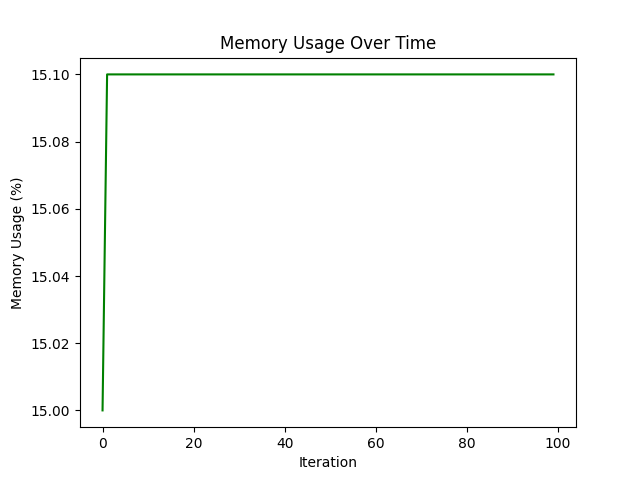}\label{fig:gesture_memory_bbai64}} \\
\vspace{-0.3cm}
\subfloat[Memory Usage for Gesture Classification (RPI4)]{\includegraphics[width=0.32\textwidth]{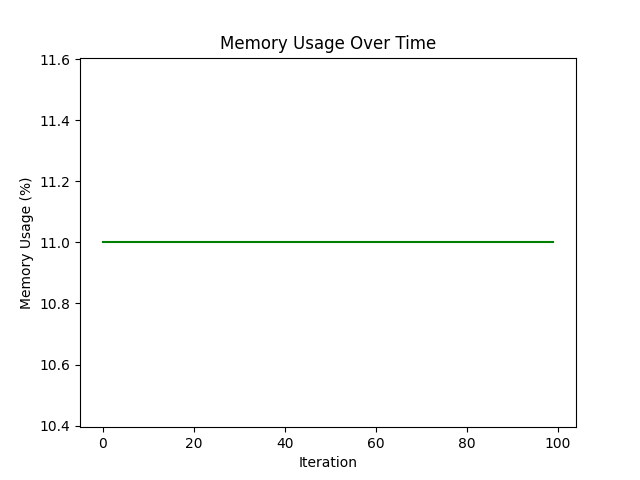}\label{fig:gesture_memory_rpi4}}
\hspace{-0.1cm}
\subfloat[CPU Usage for MobileNet V2 (BBAI64)]{\includegraphics[width=0.3\textwidth]{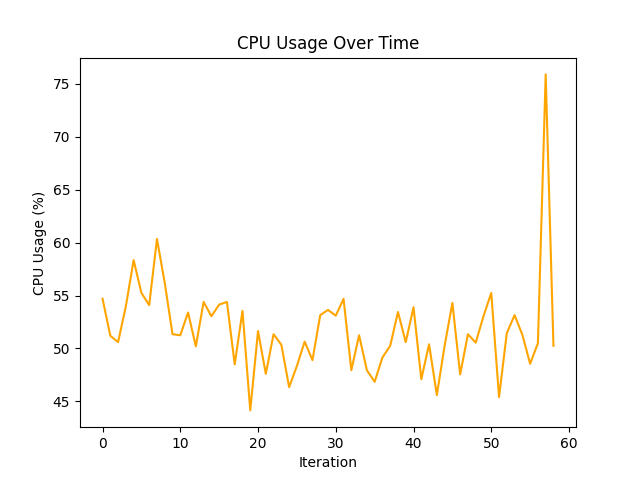}\label{fig:mn_cpu_usage_bbai64}}
\hspace{-0.1cm}
\subfloat[CPU Usage for MobileNet V2 (RPI4)]{\includegraphics[width=0.3\textwidth]{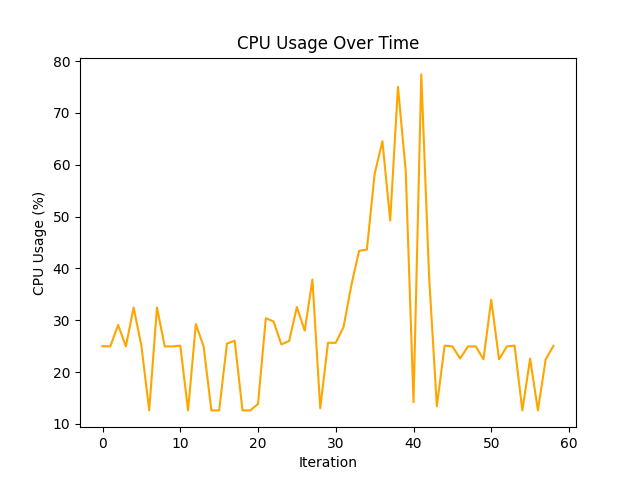}\label{fig:mn_cpu_usage_rpi4}}
\vspace{-0.2cm}
\caption{Resource Utilization Trends for Gesture Classification and MobileNet V2 Across Platforms.}
\label{fig:resource_util_comparison}
\end{figure*}

\textbf{RQ2: What are the trade-offs in resource utilization?}
Table~\ref{tab:resource_utilization} and Figure~\ref{fig:resource_util_comparison} show CPU and memory usage across platforms. \textit{Raspberry Pi 4 demonstrated significantly better resource efficiency across all models.}

\begin{table}[h]
\centering
\caption{Confidence Scores Across Platforms}
\begin{tabular}{|l|c|c|}
\hline
\textbf{Model}         & \textbf{BeagleBone AI64} & \textbf{Raspberry Pi 4} \\ \hline
Gesture Classification & 0.98                    & 0.98                   \\ \hline
Keyword Spotting       & 0.99                   & 0.99                   \\ \hline
MobileNet V2           & 17.16                   & 17.16                   \\ \hline
\end{tabular}
\label{tab:confidence_scores}
\end{table}

\begin{figure*}[ht]
\centering
{
    \includegraphics[width=0.3\textwidth]{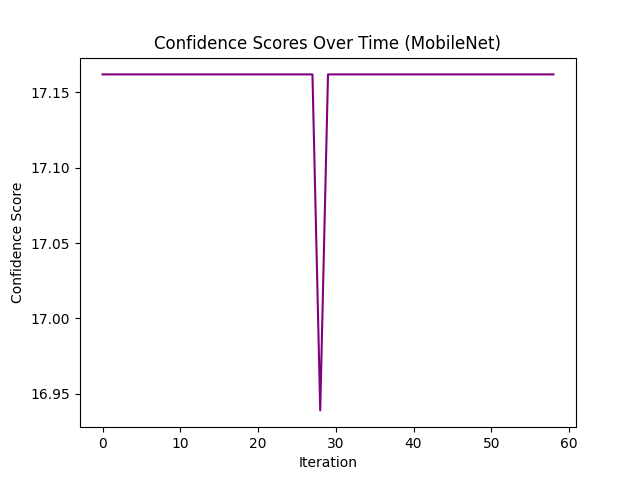}
    \label{fig:mn_confidence_scores_rpi4}}
\hfill
{
    \includegraphics[width=0.3\textwidth]{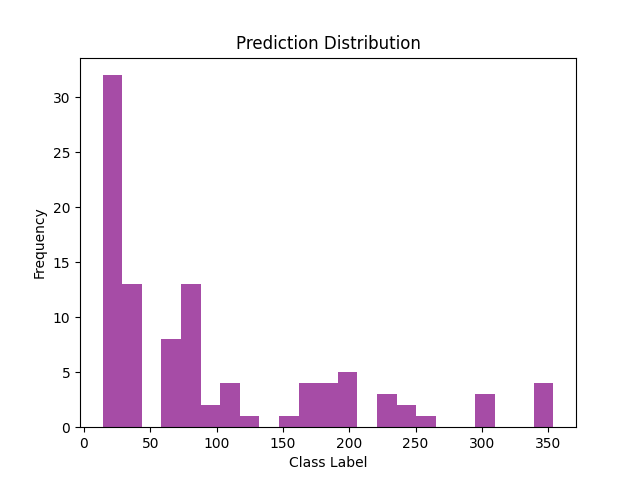}
    \label{fig:gesture_prediction_distribution_bbai64}}
\hfill
{
    \includegraphics[width=0.3\textwidth]{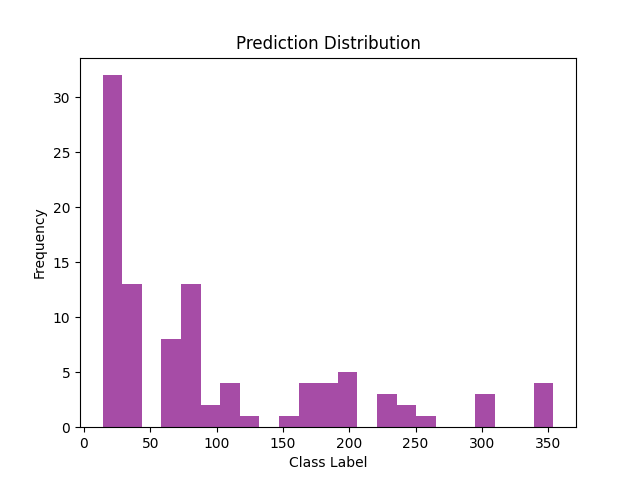}
    \label{fig:gesture_prediction_distribution_rpi4}}
\caption{Confidence scores for MobileNet V2 RPI4 (left) and prediction distribution for Gesture Classification for BBAI64 (center) and RPI64 (right).}
\label{fig:prediction_stability}
\end{figure*}

\textbf{RQ3: How stable are the predictions across iterations?} Prediction confidence scores across platforms were highly consistent, as shown in Table~\ref{tab:confidence_scores} and Figure~\ref{fig:prediction_stability}.


\textbf{Overall Comparison and Insights.}
Next, we present the results to provide a comprehensive comparison of Gesture Classification, Keyword Spotting, and MobileNet V2 models on BeagleBone AI64 and Raspberry Pi 4 platforms.

\begin{itemize}
    \item \textbf{Latency:} Raspberry Pi 4 achieved lower latency across all models, making it ideal for real-time applications $(9.49ms\mapsto1.76ms)$ for gesture classification, $(.74ms\mapsto.16ms)$ for keyword spotting, and $(2125.04\mapsto513.60ms)$ for MobileNet V2.

\tcbset{enhanced,
boxrule=0pt,frame hidden,
borderline west={4pt}{0pt}{blue},
top=0pt,bottom=0pt,
colback=blue!5!white,
sharp corners}
\begin{tcolorbox}
\textbf{Insight 1}: \textit{Raspberry Pi 4} achieved lower latency across all models.
\end{tcolorbox}

    \item \textbf{Resource Utilization:} Raspberry Pi 4 used significantly less CPU and memory, favoring battery-operated use cases. Specifically, $\sim23\%$ less CPU and $\sim6\%$ less memory consumption, respectively, than BeagleBone.

    \tcbset{enhanced,
boxrule=0pt,frame hidden,
borderline west={4pt}{0pt}{red},
top=0pt,bottom=0pt,
colback=red!5!white,
sharp corners}
\begin{tcolorbox}
\textbf{Insight 2}: Raspberry Pi 4 used significantly less CPU and memory.
\end{tcolorbox}

    \item \textbf{Prediction Stability:} Both platforms showed stable and identical prediction confidence for all the models.
\end{itemize}

\tcbset{enhanced,
boxrule=0pt,frame hidden,
borderline west={4pt}{0pt}{blue},
top=0pt,bottom=0pt,
colback=blue!5!white,
sharp corners}
\begin{tcolorbox}
\textbf{Overall Findings}: \textit{Raspberry Pi 4} is better suited for applications requiring low latency and efficient resource utilization, such as lightweight AI tasks or battery-constrained environments.
\end{tcolorbox}

\begin{figure*}[h]
\centering
{
    \includegraphics[width=0.3\textwidth]{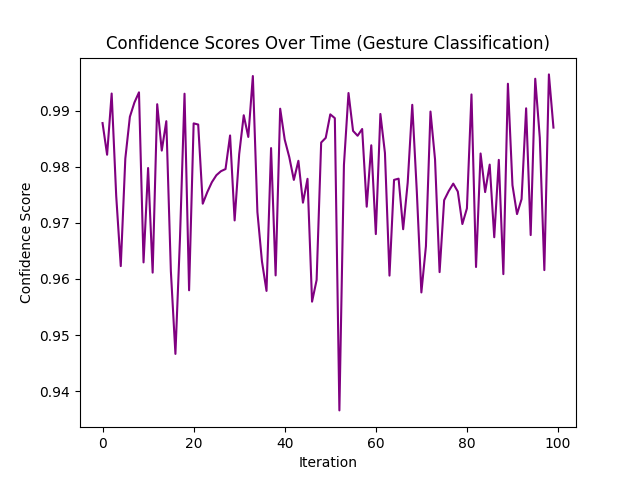}
    \label{fig:gesture_confidence_bbai64}}
\hfill
{
    \includegraphics[width=0.3\textwidth]{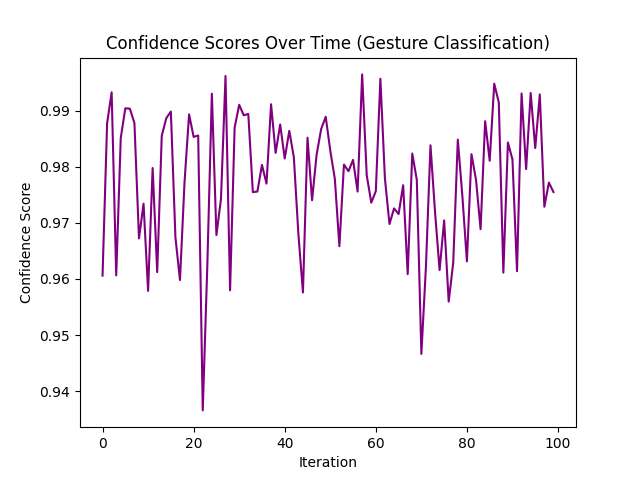}
    \label{fig:gesture_confidence_rpi4}}
\hfill
{
    \includegraphics[width=0.3\textwidth]{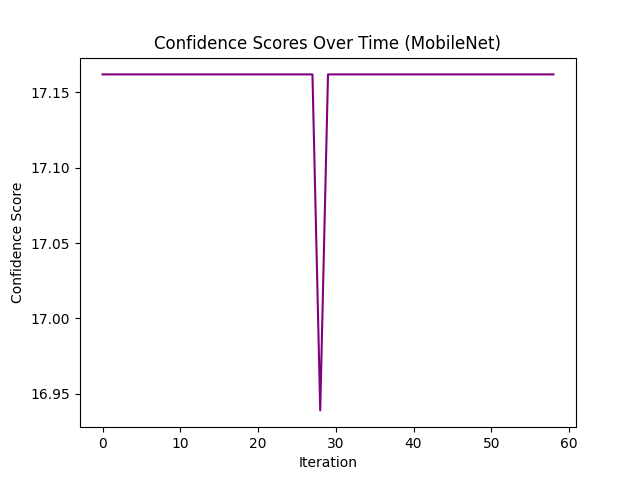}
    \label{fig:mn_confidence_scores_bbai64}}
\caption{Confidence scores across platforms for gesture classification on BBAI64 (left) and RPI4 (center), and MobileNet V2 BBAI64, (right).}
\label{fig:confidence_scores}
\end{figure*}

\section{Limitations}
While this study provides a robust foundation for benchmarking TinyML models, it has several limitations. First, the scope of the evaluation was restricted to three models and two platforms, which may not fully capture the variability of real-world TinyML deployments. Second, the framework currently omits critical metrics such as power consumption and thermal efficiency, which are vital for battery-operated and energy-constrained devices. Third, the datasets used for benchmarking were representative but limited in scale, potentially overlooking edge cases that could arise in larger and more diverse datasets. Finally, the framework assumes uniform system conditions, which might not account for variations in environmental or operational factors during deployment.

\section{Conclusion}
We present \textsc{PICO-tinyML-benchmark}, a modular framework for benchmarking TinyML models with a focus on latency, resource usage, and prediction stability under real-time conditions. We provide actionable insights for optimizing deployments across a range of embedded platforms.

In line with benchmarking efforts such as MLPerf Tiny~\cite{mlperf_tiny}, which emphasize inference accuracy, we complement existing tools by targeting system-level performance. 


\section{Future Work}

Future enhancements to \textsc{PICO-tinyML-benchmark} will focus on broadening its applicability and usability across diverse embedded systems:

\begin{itemize}
    \item \textbf{Platform Support:} Extend benchmarking to include microcontrollers and ultra-low-power processors for coverage of highly constrained environments.
    
    \item \textbf{Expanded Metrics:} Incorporate energy efficiency and thermal profiling to provide a more comprehensive evaluation of TinyML performance.
    
    \item \textbf{Model Diversity:} Add support for new model types and tasks such as object detection, time-series analysis, and multi-modal inference.
    
    \item \textbf{Usability Improvements:} Enable automated deployment pipeline integration and real-time visualization to streamline developer workflows.
    
    \item \textbf{Optimization Techniques:} Investigate dynamic quantization and adaptive resource allocation to further optimize performance under resource constraints.
\end{itemize}


\appendix
\section*{Metric Descriptions}
\label{appendix:metrics}

\begin{itemize}
    \item \textbf{Inference Latency:} The time (in milliseconds) taken to complete a single inference. Measured across 100 iterations to analyze consistency.
    \item \textbf{CPU Utilization:} The average CPU usage during inference, indicating system load and energy demand.
    \item \textbf{Memory Utilization:} The percentage of memory used before and after inference to evaluate memory efficiency.
    \item \textbf{Prediction Confidence Scores:} Probability scores or outputs associated with predicted labels to assess model stability across iterations.
\end{itemize}


\end{document}